\documentclass[a4paper,11pt]{article}
\pdfoutput=1 
\usepackage{jheppub}
\usepackage{amsfonts}
\usepackage{bm}
\usepackage{caption}
\usepackage{subcaption}
\usepackage{amsmath}
\usepackage{slashed} 
\usepackage{array}
\usepackage{longtable}
\usepackage[T1]{fontenc} 
\usepackage[usenames,dvipsnames]{xcolor}
\newcommand{\ba}{\begin{array}}
\newcommand{\ea}{\end{array}}
\newcommand{\be}{\begin{equation}}
\newcommand{\ee}{\end{equation}}
\newcommand{\beqa}{\begin{eqnarray}}
\newcommand{\eeqa}{\end{eqnarray}}
\def\321{$SU(3)\times SU(2)\times U(1)$}

\hypersetup{
  colorlinks=true,
  citecolor=blue,
  linkcolor=blue,
  urlcolor=blue}
 
\definecolor{X575}{rgb}{0.05, 0.7, 0.05}

\title{Phenomenological study of extended seesaw model for light sterile neutrino}

\author{Newton Nath$^{a,b}$,}
\author{Monojit Ghosh$^{c}$,}
\author{Srubabati Goswami$^{a}$ and} 
\author{Shivani Gupta$^{d}$}

\affiliation{$^a$Physical Research Laboratory, Navarangpura, Ahmedabad 380 009, India.}
\affiliation{$^b$Indian Institute of Technology, Gandhinagar, Ahmedabad--382424, India.}
\affiliation{$^c$Department of Physics, Tokyo Metropolitan University, Hachioji, Tokyo 192-0397, Japan.}
\affiliation{$^d$Center of Excellence for Particle Physics (CoEPP), University of Adelaide, Adelaide SA 5005, Australia.}

\emailAdd{newton@prl.res.in}
\emailAdd{monojit@tmu.ac.jp}
\emailAdd{sruba@prl.res.in} 
\emailAdd{shivani.gupta@adelaide.edu.au}

\abstract{We study the zero textures of the Yukawa matrices in the minimal extended type-I seesaw (MES) model which can give rise to $\sim$ eV scale sterile neutrinos. 
In this model, three right handed neutrinos and one extra singlet $S$ are  added to generate
a light sterile neutrino. The light neutrino mass matrix for the active 
neutrinos, 
$ m_{\nu}$, depends on the Dirac neutrino mass matrix ($ M_{D} $), 
Majorana neutrino mass matrix ($ M_{R} $) and the mass matrix
($ M_{S} $) coupling the right handed neutrinos and the singlet.  
The model predicts one of the light neutrino masses to vanish. 
We systematically investigate the  zero textures in $ M_{D} $
and observe that maximum five zeros in $ M_{D} $  can lead to viable zero 
textures in $ m_{\nu} $.
For this study we consider four different forms for  $ M_R $ (one diagonal and three off diagonal) and  two different forms of $(M_{S})$ containing  one zero.
Remarkably  we obtain only two  allowed forms of $ m_{\nu} $  
($m_{e\tau} = 0 $ and $m_{\tau\tau}=0$) having  
inverted hierarchical mass spectrum. 
We re-analyze the phenomenological
implications of these two allowed  
textures of $m_\nu$ in the light of recent neutrino oscillation data. 
In the context of the MES model, we also  express the low energy mass matrix,
the mass of the sterile neutrino  and the active-sterile mixing in terms of the
parameters of the allowed Yukawa matrices. 
The MES model leads to some extra correlations which disallow 
some of the Yukawa textures obtained earlier, 
even though they give allowed one-zero forms of $m_\nu$. 
We show that the allowed textures in our study can be realized in a simple way in a model
based on MES mechanism with a discrete Abelian flavor symmetry group $Z_8 \times Z_2$.
}

\begin{document}
\preprint{ADP-16 - 34/T990} 
\maketitle

\section{Introduction}

Neutrino oscillation experiments have established the fact that neutrinos have tiny mass and  
they change from one flavor to another during their propagation. 
This requires the Standard Model (SM) of particle physics to be extended in order to generate their masses. The standard 3-flavor neutrino oscillation scenario has six key parameters. 
These are the two mass squared differences ($\Delta m^2_{i1}, = m_i^2-m_1^2,i =2,3$ ) 
which control the oscillations of the solar and atmospheric neutrinos respectively, 
three mixing angles $\theta_{ij}$ ($ i,j = 1,2,3; i<j $) and a Dirac CP phase, $ \delta_{13} $. 
Global analysis of three flavor neutrino oscillation data from 
\cite{Gonzalez-Garcia:2015qrr,Forero:2014bxa,Capozzi:2016rtj} 
give us the best fit values and the allowed $ 3\sigma $ ranges of these
parameters. In 3-flavor paradigm, there are two more CP violating phases if neutrinos are
Majorana particles. But as Majorana phases do not appear in the neutrino oscillation probability, they are not measurable 
in the oscillation experiments. 
Apart from these phases another major unknown is the
absolute value of the neutrino mass since oscillation experiments are only sensitive 
to the mass squared differences. 
Planck data provide an upper bound on sum of neutrino masses to be 
$ \leq 0.23$ eV \cite{Ade:2015xua} at 95\% C.L. 
The sensitivity for the neutrino masses
in the upcoming Karlsruhe Tritium Neutrino experiment (KATRIN)  is 
expected to be around 
200 meV (90\% C.L.) \cite{MERTENS2015267}.

Another interesting aspect of neutrino oscillation experiments is the
search for the existence of a 
light sterile neutrino. As sterile neutrinos are SM singlets they do not
take part in the weak interactions.  
But they can mix with the active neutrinos. 
Therefore, sterile neutrinos can be probed in neutrino oscillation 
experiments. The oscillation results from LSND experiment showed the
evidence of at least one sterile neutrino
having mass in the $\sim$ eV scale
\cite{Athanassopoulos:1996jb,Athanassopoulos:1997pv,Aguilar:2001ty}. 
The latest data of MiniBooNE experiment \cite{Aguilar-Arevalo:2013pmq}
also have some overlap with the allowed regions of
the LSND experiment and hence support the existence of the sterile neutrino
hypothesis. 
The recently observed Gallium anomaly  can also be explained by the
sterile neutrino hypothesis 
\cite{Giunti:2010zu}. 
Another evidence of eV sterile neutrino comes from the reactor
antineutrino flux studies. 
This shows the deficit in the observed and predicted event rate of 
electron antineutrino flux and the ratio is $ 0.943 \pm 0.023 $ at 98.6\% C.L. \cite{Mention:2011rk}. Recent analysis of the Planck data shows the possibility
of light sterile neutrino in the eV scale if one deviates slightly from the
base $\Lambda$CDM model \cite{Ade:2015xua}. 
In short, the scenario with a light sterile neutrino 
is quite riveting at present and many future experiments are proposed to confirm/falsify this \cite{Abazajian:2012ys}.
Although it is possible to have a better fit of neutrino oscillation data
with more than one light sterile neutrino
\cite{Kopp:2011qd,Conrad:2012qt,Giunti:2011gz}, the 3+1 scheme i.e.,
three active neutrinos and one sterile neutrino in the sub-eV and eV scale
respectively, is considered to be minimal. There are three different ways to
add sterile neutrino in SM mass patterns and these are, (i) 3+1 scheme in which three active
neutrinos are of sub-eV scale and  sterile neutrino  is of eV scale \cite{GomezCadenas:1995sj,Goswami:1995yq}, (ii) 2+2 scheme in which two different pairs of neutrino mass states differ by eV$^{2} $ but this scheme was disfavored
by solar and atmospheric data \cite{Maltoni:2002ni}, and (iii) 1+3 scheme in which three active neutrinos are in eV
scale and sterile neutrino is lighter than active neutrinos. This scenario is however disfavored from cosmology \cite{Hamann:2010bk,Giusarma:2011ex}.
Hence, we focus on the 3+1 scenario in our study. 

Flavor symmetry models giving rise to eV sterile neutrinos have been studied in
the literature \cite{Chun:1995bb,Barry:2011wb,Chen:2011ai}.
These models might require modifications to usual seesaw framework \cite{deGouvea:2006gz,Dev:2012bd}. In the explicit seesaw models the eV scale sterile
neutrinos with their mass suppressed
by Froggatt - Nielsen mechanism can be naturally accommodated in non Abelian
$A_4$ flavor symmetry \cite{Barry:2011wb,Zhang:2011vh,Heeck:2012bz}. 
$S_3$ bimodel or schizophrenic
models for light sterile neutrinos are also widely studied 
\cite{Allahverdi:2010us,Machado:2010ui}.
In order to have a theoretical understanding of the origin of eV 
sterile neutrino as well as admixtures
between sterile and active neutrinos, 
the authors of Refs. \cite{Barry:2011wb,Zhang:2011vh,Heeck:2012bz} have studied
an extension to the canonical type-I seesaw model. This model is known as ``minimal extended type - I seesaw" (MES) model.  
In the MES model a fermion singlet, $S$, is added along with three right handed neutrinos. This extension results
into an eV scale sterile neutrino naturally, without imposing tiny mass scale or Yukawa term for this neutrino. 

In this paper, for the first time we study the various possible textures of the Dirac neutrino mass matrix,
$M_{D} $, Majorana neutrino mass matrix, $ M_{R} $ and the mass matrix $ M_{S} $ that originate from the
Yukawa interaction between right handed neutrinos with the gauge singlet within the framework of MES model and classify the allowed possibilities.
Several papers have studied the consequences of imposing zeros in 
the neutrino mass matrix in standard three neutrino 
\cite{Dev:2006qe,Xing:2002ta,Xing:2002ap,Desai:2002sz,Dev:2007fs,Dev:2006xu,Kumar:2011vf,Fritzsch:2011qv,
Meloni:2012sx,Ludl:2011vv,Grimus:2012zm} and the 3+1 framework \cite{Ghosh:2012pw,Ghosh:2013nya,Zhang:2013mb,Nath:2015emg,Borah:2016xkc}. 
The more natural study would be to explore the zeros in the
Yukawa matrices that appear in
the Lagrangian rather than light neutrino mass matrix, $ m_{\nu} $. 
It has been noted by many authors \cite{Branco:2007nb,Goswami:2008rt,Goswami:2008uv,Choubey:2008tb,Lavoura:2015wwa}
that the zeros of the Dirac neutrino mass matrix $M_D$ and the right handed Majorana mass matrix
$M_R$ are the progenitors of zeros in the effective Majorana mass matrix
$m_{\nu}$ through type - I seesaw mechanism.
We also seek extra correlations connecting the parameters of the active 
and sterile sector which can put further constraints on the 
allowed possibilities. 
This motivates us to look for zeros in various neutrino mass matrices in the 
MES model which can lead to viable texture zeros in neutrino mass matrix.

We classify different structures of $M_D$, $M_R$ and $M_S$ that can give allowed textures for the 
light neutrino mass matrix $m_\nu$. 
Interestingly  the only allowed form of $m_\nu$ that 
we obtain  are the two  one zero textures  
-- namely $ m_{e \tau} = 0$ and $ m_{\tau \tau} = 0$ which are 
phenomenologically allowed and have the inverted hierarchical mass spectrum. 
For a $m_\nu$ originating from ordinary seesaw mechanism both 
these textures are viable. However, in the MES model, because of 
extra correlations connecting active and sterile sector, not all Yukawa
matrices that give $ m_{e \tau} = 0$ or $ m_{\tau \tau} = 0$ for $m_\nu$ 
are allowed. We study these additional correlations and tabulate the 
allowed textures. 
 We also include a discussion on the impact of NLO corrections in this model.
In this context it is also important to study the origin of zero textures.
Here, we show that it is possible to obtain various zero entries in lepton mass matrices
with an Abelian discrete symmetry group $Z_8 \times Z_2$. 
An alternative approach to obtain lepton mixing is discussed in \cite{Fonseca:2014koa} by considering non-Abelian symmetry group. 
We follow the method discussed in \cite{Grimus:2004hf} to obtain Abelian discrete symmetry group which
can generate viable zero textures in $m_{\nu}$. Their method is based on type - I seesaw and we extend it to apply on MES model.

The paper is organized in the following manner.
In the next section a brief review of the MES model is given. 
In Section \ref{sec3} and its subsections we list the various forms of $M_D$, 
$M_R$ and $M_S$ that lead to viable textures in $m_\nu$.
In Section \ref{sec4} we discuss the implication of the allowed forms of
one zero textures in $m_{\nu}$ obtained in Section \ref{sec3}. 
The following Section \ref{sec5} discusses the results obtained from the comparison of low energy
and high energy neutrino mass matrices and the extra correlations connecting 
active and sterile sector. 
Symmetry realizations for the allowed zero textures are discussed in Section \ref{sec6}.
The summary of our findings and conclusions are presented in Section \ref{sec7}. 

\section{Minimal extended type I seesaw mechanism}
\label{sec2}

In this section we describe the basic structure of MES model. 
Here, the fermion content of the SM is extended by three right handed neutrinos 
together with a gauge singlet field $S$. 
One can get a natural eV-scale sterile neutrino without inserting  any small
Yukawa coupling in
this model \cite{Barry:2011wb, Zhang:2011vh}. 
The Lagrangian containing the  neutrino masses is given by,
\begin{equation}
-\mathcal{L_{M}} = \overline{\nu_L}M_D \nu_R + \overline{S^c}M_S \nu_R +\frac{1}{2}\overline{\nu_R^c}M_R\nu_R +h.c..
\end{equation}
Here, $M_{D}, M_{R}$ are the ($ 3\times3 $) Dirac and Majorana mass matrices respectively and $ M_{S} $ is a ($ 1\times3 $)  coupling matrix  between right handed neutrinos with the gauge singlet.
In the basis ($ \nu_{L},\nu^{c}_{R},S^{c} $), the ($ 7\times7 $) neutrino mass matrix can be expressed as,
\begin{equation}\label{seesaw}
M_{\nu}^{7\times7} = \left(
\begin{array}{ccc}
0 & M_{D} & 0 \\
M^{T}_{D} & M_R & M^{T}_{S} \\
0 & M_S & 0
\end{array}
\right).
\end{equation}
Considering the hierarchical mass spectrum of these mass matrices i.e. $ M_R \gg M_S > M_D  $, in analogy of type - I seesaw,
the right handed neutrinos are much heavier compared to the electroweak scale and thus they will decouple at the low scale. 
Therefore, Eq.(\ref{seesaw}) can be block diagonalized using seesaw mechanism and the effective
neutrino mass matrix in the basis ($ \nu_{L},S^{c} $) can be written as,
\begin{equation}
M_{\nu}^{4\times4} = - \left(
\begin{array}{cc}
M_DM_R^{-1}M_D^T & M_DM_R^{-1}M_S^T  \\
M_S(M_R^{-1})^TM_D^T & M_SM_R^{-1}M_S^T 
\end{array}
\right).
\label{mnus}
\end{equation}
Note that the rank of $ M_{\nu}^{4\times4} $ is three (see \cite{Zhang:2011vh}) and hence one of the light neutrino remains massless.

Considering the case that $ M_S > M_D $, one can apply seesaw approximation 
once again on Eq.(\ref{mnus})
to obtain the active neutrino mass matrix as\footnote{Note that RHS of Eq.(\ref{mnu}) does not vanish
since $ (M_S)_{1\times3} $ is a vector rather than a square matrix.},
\begin{equation}
m^{{3\times3}}_{\nu} \simeq M_DM_R^{-1}M_S^T (M_SM_R^{-1}M_S^T)^{-1} M_S(M_R^{-1})^TM_D^T - M_DM_R^{-1}M_D^T,
\label{mnu}
\end{equation}
whereas the mass of the sterile neutrino is given by,
\begin{equation}
m_{s} \simeq - M_SM_R^{-1}M_S^T.
\label{mss}
\end{equation}
Note that the 
zero textures of fermion mass matrices in the context of type - I seesaw
mechanism studied in 
\cite{Branco:2007nb,Goswami:2008rt,Goswami:2008uv,Lavoura:2015wwa}, 
leading to viable texture zeros in $ m^{{3\times3}}_{\nu} $ can be different 
from that of MES model because of the presence of the first term of 
Eq.(\ref{mnu}).
The active-sterile neutrino mixing matrix is given by,
\begin{equation}
V \simeq \left(
\begin{array}{cc}
(1- \frac{1}{2}RR^{\dagger})U^{\prime} & R \\
-R^{\dagger}U^{\prime} & 1- \frac{1}{2}R^{\dagger}R
\end{array}
\right),
\end{equation}
where $ R_{3\times1} $ governs the strength of active-sterile mixing and can be expressed as,
\begin{equation}
  R_{3\times1} = M_DM_R^{-1}M_S^T (M_SM_R^{-1}M_S^T)^{-1}.
  \label{rr}
 \end{equation}
Essentially, $ R_{3\times1}  = (V_{e 4},V_{\mu 4},0)^T$ 
is suppressed by the ratio $ \mathcal{O}(M_D)/\mathcal{O}(M_S) $. Additionally in our formalism we assume $ |V_{\tau 4}| = 0 $,
which is allowed by the current active sterile neutrino mixing data.

\begin{figure}
 \begin{center}
 \includegraphics[scale=0.42]{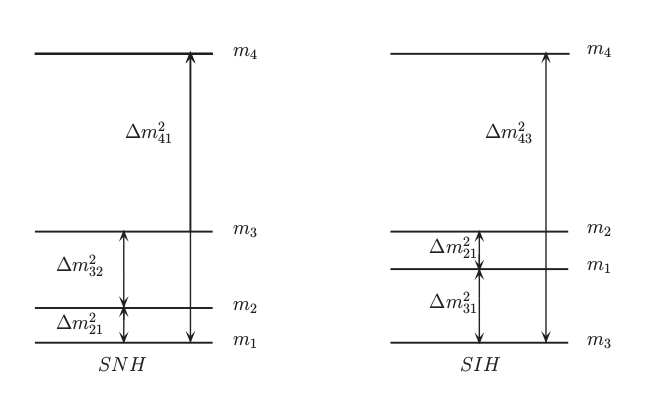}
\end{center}
\vspace{-0.5cm}
\caption{Allowed  mass spectrum in 3+1 scheme for normal (SNH) and inverted (SIH) mass hierarchy.}
\label{fig1}
 \end{figure}

As the sterile neutrino mass ($ \sim $ eV) is heavier than active neutrinos, 
therefore, the mass pattern in the active sector can be arranged in two 
different ways.
We denote 3+1 scenario as (SNH) when the  three active neutrinos follow 
normal hierarchy ($ m_1 < m_2 \ll m_3 $) and the second choice is (SIH)  
when the three active neutrinos follow inverted hierarchy 
$ m_3 \ll m_1 \approx m_2 $) as shown in Fig(\ref{fig1}). 
These masses can be expressed in terms of  the mass squared differences
obtained from oscillation experiments as given in Table(\ref{mass}).
\begin{table}[!h]
\centering
\begin{tabular}{|c|c|c|}
\hline
 & SNH & SIH \\ 
\hline
$m_1$ & 0 & $ \sqrt{\Delta m^{2}_{31}} $  \\
\hline
$m_2$ & $ \sqrt{\Delta m^{2}_{21}} $ & $ \sqrt{\Delta m^{2}_{21}+\Delta m^{2}_{31}} $  \\
\hline
$m_3$ & $ \sqrt{\Delta m^{2}_{21}+\Delta m^{2}_{32}} $ & 0 \\
\hline
$m_4$ & $ \sqrt{\Delta m^{2}_{41}} $ & $ \sqrt{\Delta m^{2}_{43}} $ \\
\hline
\end{tabular}
\begin{center}
\caption{Neutrino mass spectrum for normal and inverted hierarchies. $\Delta m^2_{12}$,  $\Delta m^{2}_{31}$ ($\Delta m^{2}_{32}$) are the solar and atmospheric mass squared differences and  $\Delta m^{2}_{41}$ ($\Delta m^{2}_{43}$) is the active sterile mass squared difference. The allowed ranges of these three
mass squared differences are given in Table(\ref{parameters}).
} 
\label{mass}
\end{center}
\end{table}
The best fit values along with 3$\sigma$ ranges of neutrino oscillation
parameters used in our numerical analysis are given in Table(\ref{parameters}).
\begin{table} [!h] 
\centering
\begin{tabular}{||c|cc||}
\hline
Parameter &  Best Fit  & $3\sigma$ Range \\
\hline
\hline
$\Delta m^2_{21}[10^{-5}~\mathrm{eV}^2] $  & 7.37& 6.93 -- 7.97 \\
\hline
$\Delta m^2_{31}[10^{-3}~\mathrm{eV}^2] $ (NH) & 2.50& 2.37 -- 2.63 \\
$\Delta m^2_{31}[10^{-3}~\mathrm{eV}^2] $ (IH) & 2.46 & 2.33 -- 2.60 \\
\hline
$\sin^2 \theta_{12}/10^{-1}$  & 2.97 & 2.50 -- 3.54 \\
\hline
$\sin^2 \theta_{13}/10^{-2}$ (NH) & 2.14 & 1.85 -- 2.46 \\
$\sin^2 \theta_{13}/10^{-2}$ (IH) & 2.18 & 1.86 -- 2.48 \\
\hline
$\sin^2 \theta_{23}/10^{-1}$ (NH)  & 4.37  &  3.79 -- 6.16 \\
$\sin^2 \theta_{23}/10^{-1}$ (IH)  & 5.69  &  3.83 -- 6.37 \\
\hline
$\delta_{13}/ \pi$ (NH) & 1.35 & 0 -- 2 \\
$\delta_{13}/ \pi$ (IH) & 1.32 & 0 --2 \\
\hline
$R_{\nu}$(NH) & 0.0295 & 0.0263 -- 0.0336\\
$R_{\nu}$(IH) & 0.0299 & 0.0266 -- 0.0342\\
\hline
$\rm \Delta m_{LSND}^2(\rm \Delta m_{41}^2 \rm or \Delta m_{43}^2) ~\mathrm{eV}^2$  & 1.63 & 0.87 -- 2.04 \\
$ |V_{e 4}|^{2} $ & 0.027 & 0.012 -- 0.047 \\
$ |V_{\mu 4}|^{2} $  & 0.013 & 0.005 -- 0.03 \\
$ |V_{\tau 4}|^{2} $ &  --  & $ <$ 0.16  \\
\hline
\end{tabular}
\begin{center}
\caption{The latest best-fit and $3 \sigma$ ranges of active $ \nu $ oscillation parameters from \cite{Capozzi:2016rtj}. 
The current constraints on sterile neutrino parameters are from the global analysis \cite{Gariazzo:2015rra,Giunti_neutrino_2016,schwetz}. Here $R_{\nu}$ is
the solar to atmospheric mass squared difference ratio.}
\label{parameters}
\end{center}
\end{table}
In the next section we systematically explore the various zero texture structures of $ M_{D}$, $ M_{R}$ and $ M_{S}$
which can give rise to viable zero textures of $ m^{{3\times3}}_{\nu} $.

\section{Formalism}
\label{sec3}

In our formalism, the charge lepton mass matrix, $ M_{l} $, is considered 
to be diagonal.
For the right handed Majorana neutrino mass matrix, 
we  consider  four different structures:\\
(i) Diagonal $ M_{R}$ having three zeros i.e.,
\begin{equation}
M_R = \left(
\begin{array}{ccc}
r_1 & 0 & 0\\
0 & r_2 & 0 \\
0 & 0 & r_{3}
\end{array}
\right)
\label{diag_mr}
\end{equation}
\\
(ii) non-diagonal minimal form of $ M_{R}$ having four zeros with  Det $ M_{R} \neq 0$
i.e., 
\begin{equation}
M_R =  
\left(
\begin{array}{ccc}
0 & r_2 & 0\\
r_2 & 0 & 0 \\
0 & 0 & r_1
\end{array}
\right);~~~
\left(
\begin{array}{ccc}
0 & 0 & r_2\\
0 & r_1 & 0 \\
r_2 & 0 & 0
\end{array}
\right);~~~
\left(
\begin{array}{ccc}
r_1 & 0 & 0\\
0 & 0 & r_2 \\
0 & r_2 & 0
\end{array}
\right).
\label{mr_nondiag}
\end{equation}
These three non-diagonal forms of $M_R$ correspond to
 $ L_e - L_\mu $, $ L_e - L_\tau $ and $ L_\mu - L_\tau $ flavor symmetry respectively. Such forms of $M_R$ in the context of zero textures in type-I seesaw model have been considered for instance in \cite{Goswami:2009bd}.
$ M_S =(s_{1},s_{2},s_{3})$ being a $ 1\times3 $ matrix can have one
zero or two zeros. In \cite{Zhang:2011vh},  an $ A_4 $ based model was considered with 2 zeros in  $ M_S $ and 3 zeros in $ M_D $ to obtain the  $m^{3\times3}_\nu$ as given by Eq. (\ref{mnu}). But, in our analysis we find that mass matrices with 5 zeros in $ M_D $ and   two zeros in $ M_S $ do not lead to any viable textures in $m_\nu$. 
The only allowed possibility therefore is one zero in $M_S$ result in  three 
possible structures.  
We find that the maximum number of zeros of $M_D$ that can give phenomenologically allowed zero textures in $m_\nu$ is five. 
The possible combinations of $ M_{D}$, $M_R$ and 
$M_S$ that lead to
phenomenologically viable textures of $m_{\nu}$ are discussed 
in the following subsections.

\subsection{5 zeros in $ M_D $ and diagonal $M_R$}

First let us assume $M_R$ to be diagonal. As $ M_D $ is a non-symmetric $ 3\times3 $ matrix, 5 zeros can be arranged in $ ^9C_5 =126$ ways. Thus considering 126 cases of $ M_D $ together with 3 cases of $ M_S $ and 1 case of $M_R$, we obtain total 378 possible structures of $ m_{\nu} $. Out of all possible combinations of these matrices the only allowed texture that we  obtain is the one zero texture in $m_\nu$ with $ m_{e \tau}=0$.
Here, we have three possible forms of $ M_S $ and these are,
\begin{equation}
M^{(1)}_S =(0,s_{2},s_{3}),~M^{(2)}_S =(s_{1},0,s_{3}),~ {\rm and} ~ M^{(3)}_S =(s_{1},s_2,0).
\end{equation}
The various forms of $ M_D $  which lead to viable texture $  m_{e \tau} =0  $ are presented below: 
\begin{equation}
M^{(1)}_S ,M^{(1)}_{D}= \left(
\begin{array}{ccc}
0 & 0 & a_{3}\\
b_{1} & 0 & b_{3} \\
c_{1} & 0 & 0
\end{array}
\right),M^{(2)}_{D}=
\left(
\begin{array}{ccc}
0 & a_{2} & 0\\
b_{1} & 0 & b_{3} \\
c_{1} & 0 & 0
\end{array}
\right),M^{(3)}_{D}=M^{(1)}_{D}Z_{23},M^{(4)}_{D}=M^{(2)}_{D}Z_{23} .
\label{etau_1}
\end{equation}
\begin{equation}
M^{(2)}_S , M^{(5)}_{D} = \left(
\begin{array}{ccc}
0 & 0 & a_{3}\\
0 & b_{2} & b_{3} \\
0 & c_{2} & 0
\end{array}
\right),M^{(6)}_{D} =
\left(
\begin{array}{ccc}
a_{1} & 0 & 0\\
0 & b_{2} & b_{3} \\
0 & c_{2} & 0
\end{array}
\right),M^{(7)}_{D}=M^{(5)}_{D}Z_{13},M^{(8)}_{D}=M^{(6)}_{D}Z_{13}.
\label{etau_2}
\end{equation}
\begin{equation}
M^{(3)}_S , M^{(9)}_{D} = \left(
\begin{array}{ccc}
a_{1} & 0 & 0\\
0 & b_{2} & b_{3} \\
0 & 0 & c_{3}
\end{array}
\right),M^{(10)}_{D} =
\left(
\begin{array}{ccc}
0 & a_{2} & 0\\
0 & b_{2} & b_{3} \\
0 & 0 & c_{3}
\end{array}
\right),M^{(11)}_{D}=M^{(9)}_{D}Z_{12},M^{(12)}_{D}=M^{(10)}_{D}Z_{12}.
\label{etau_3}
\end{equation}
Here, $ Z_{12} $, $ Z_{13} $ and $ Z_{23} $ are the permutation matrices that exchange first and second columns,
first and third columns and second and third columns respectively.
Therefore, we observe that out of 126 cases only 12 above forms of $ M^{(i)}_{D},i=1-12 $ give the allowed texture $ m_{e \tau}=0$ of $m_\nu$ when $M_R$ is diagonal 

\subsection{5 zeros in $ M_D $ and non-diagonal $M_R$ corresponding to $ L_e -L_\mu $ flavor symmetry}

The form of $M_R$ that we consider here corresponds to flavor symmetry $ L_e -L_\mu $ as given in Eq.(\ref{mr_nondiag}). 
Among the 378 possibilities we obtain two allowed one zero textures of $m_\nu$, namely $ m_{e \tau} = 0 $ and $ m_{\tau \tau} = 0$. 
We observe that out of total 126 forms of $ M_{D} $, only four structures give rise to $ m_{e \tau}=0$ while eight
structures give rise to $ m_{\tau \tau}=0$. We list them below:
\subsubsection{Textures leading to $ m_{e \tau}=0$}
\begin{equation}
M^{(3)}_S ,M^{(13)}_{D}= \left(
\begin{array}{ccc}
a_1 & 0 & 0\\
0 & b_2 & b_3 \\
0 & 0 & c_3
\end{array}
\right),M^{(14)}_{D}=
\left(
\begin{array}{ccc}
0 & a_2 & 0\\
0 & b_2 & b_3 \\
0 & 0 & c_3
\end{array}
\right),M^{(15)}_{D}=M^{(13)}_{D}Z_{12},M^{(16)}_{D}=M^{(14)}_{D}Z_{12}.
\label{eq:e_tau_1_fv1}
\end{equation}
\subsubsection{Textures leading to $m_{\tau \tau}=0$}
\begin{eqnarray} \label{eq:md18_mtautau}
M^{(1)}_S  ,&M^{(17)}_{D}= 
\left(
\begin{array}{ccc}
a_1 & a_2 & 0\\
b_1 & 0 & 0 \\
0 & c_2 & 0
\end{array}
\right),M^{(18)}_{D}=\left(
\begin{array}{ccc}
a_1 & 0 & a_{3}\\
b_1 & 0 & 0 \\
0 & c_2 & 0
\end{array}
\right),\\ \nonumber & M^{(19)}_{D}=
\left(
\begin{array}{ccc}
a_1 & 0 & 0\\
b_1 & b_2 & 0 \\
0 & c_2 & 0
\end{array}
\right)
,M^{(20)}_{D}= \left(
\begin{array}{ccc}
a_1 & 0 & 0\\
b_1 & 0 & b_3 \\
0 & c_2 & 0
\end{array}
\right).
\label{tau_tau_1}
\end{eqnarray}
\begin{eqnarray}
M^{(2)}_S  ,& M^{(21)}_{D}= \left(
\begin{array}{ccc}
0 & a_2 & a_{3}\\
0 & b_2 & 0 \\
c_1 & 0 & 0
\end{array}
\right), M^{(22)}_{D}= \left(
\begin{array}{ccc}
a_1 & a_2 & 0\\
0 & b_2 & 0 \\
c_1 & 0 & 0
\end{array}
\right),\\ \nonumber &
M^{(23)}_{D}= \left( \begin{array}{ccc}
0 & a_2 & 0\\
b_1 & b_2 & 0 \\
c_1 & 0 & 0
\end{array}
\right)
,M^{(24)}_{D}= \left(
\begin{array}{ccc}
0 & a_2 & 0\\
0 & b_2 & b_3 \\
c_1 & 0 & 0
\end{array}
\right).
\label{tau_tau_2}
\end{eqnarray}
\subsection{5 zeros in $ M_D $ and non-diagonal $M_R$ corresponding to $ L_e -L_\tau $ flavor symmetry}
The form of $M_R$ that we consider in this subsection
corresponds to flavor symmetry $ L_e -L_\tau $ as given in 
Eq.(\ref{mr_nondiag}). 
In this case also we observe that out of total 126 cases of $ M_{D} $, 
only four structures of $ M_{D} $ give rise to $ m_{e \tau}=0$
and eight forms of $ M_{D} $ give rise to texture $ m_{\tau \tau}=0$. 
We list them below. Note that these forms of $ M_{D} $ are different from those obtained in the earlier subsection. 
\subsubsection{Textures leading to $ m_{e \tau}=0$}
\begin{equation}
M^{(2)}_S ,M^{(25)}_{D}=\left(
\begin{array}{ccc}
0 & 0 & a_3\\
0 & b_2 & b_3 \\
0 & c_2 & 0
\end{array}
\right)
,M^{(26)}_{D}=\left(
\begin{array}{ccc}
0 & 0 & a_3\\
b_1 & b_2 & 0 \\
0 & c_2 & 0
\end{array}
\right),M^{(27)}_{D}=M^{(25)}_{D}Z_{13},M^{(28)}_{D}=M^{(26)}_{D}Z_{13}.
\label{mutau_fv2}
\end{equation}
\subsubsection{Textures leading to $ m_{\tau \tau}=0$}
\begin{eqnarray}
M^{(1)}_S ,&M^{(29)}_{D}= \left(
\begin{array}{ccc}
a_1 & a_2 & 0\\
b_1 & 0 & 0 \\
0 & 0 & c_3
\end{array}
\right), M^{(30)}_{D} =
\left(
\begin{array}{ccc}
a_1 & 0 & a_3\\
b_1 & 0 & 0 \\
0 & 0 & c_3
\end{array}
\right),\\ \nonumber &M^{(31)}_{D} =
\left(
\begin{array}{ccc}
a_1 & 0 & 0\\
b_1 & b_2 & 0 \\
0 & 0 & c_3
\end{array}
\right) 
,M^{(32)}_{D} = \left(
\begin{array}{ccc}
a_1 & 0 & 0\\
b_1 & 0 & b_3 \\
0 & 0 & c_3
\end{array}
\right).
\label{tautau_fv2}
\end{eqnarray}
\vspace{-0.5cm}
\begin{eqnarray}
M^{(3)}_S ,&M^{(33)}_{D}= \left(
\begin{array}{ccc}
0 & a_2 & a_3\\
0 & 0 & b_3 \\
c_1 & 0 & 0
\end{array}
\right),M^{(34)}_{D} =
\left(
\begin{array}{ccc}
a_1 & 0 & a_3\\
0 & 0 & b_3 \\
c_1 & 0 & 0
\end{array}
\right),\\ \nonumber &M^{(35)}_{D} =
\left(
\begin{array}{ccc}
0 & 0 & a_3\\
0 & b_2 & b_3 \\
c_1 & 0 & 0
\end{array}
\right)
,M^{(36)}_{D} = \left(
\begin{array}{ccc}
0 & 0 & a_3\\
b_1 & 0 & b_3 \\
c_1 & 0 & 0
\end{array}
\right).
\label{tautau_fv2}
\end{eqnarray}
\subsection{5 zeros in $ M_D $ and non-diagonal $M_R$ corresponding to $ L_\mu -L_\tau $ flavor symmetry}
The form of $M_R$ that we consider here corresponds to flavor symmetry $ L_\mu -L_\tau $ as given in Eq.(\ref{mr_nondiag}). 
Here also we observe that out of 126 cases of $ M_{D} $ only four structures of $ M_{D} $ give rise to texture $ M_{e \tau}=0$
and 8 forms of $ M_{D} $ give rise to texture $ M_{\tau \tau}=0$. 
But these forms of $ M_{D} $ are different from those obtained in the earlier two subsections:
\subsubsection{Structures leading to $ m_{e \tau}=0$}
\begin{equation}
M^{(1)}_S  ,M^{(37)}_{D}= \left(
\begin{array}{ccc}
0 & a_2 & 0\\
b_1 & 0 & b_3 \\
c_1 & 0 & 0
\end{array}
\right),M^{(38)}_{D}=\left(
\begin{array}{ccc}
0 & 0 & a_3\\
b_1 & 0 & b_3 \\
c_1 & 0 & 0
\end{array}
\right),M^{(39)}_{D}=M^{(37)}_{D}Z_{23},M^{(40)}_{D}=M^{(38)}_{D}Z_{23}.
\label{etau_fv3}
\end{equation}
\subsubsection{Structures leading to $ m_{\tau \tau}=0$}
\begin{eqnarray}
M^{(2)}_S  ,&M^{(41)}_{D}= \left(
\begin{array}{ccc}
a_1 & a_2 & 0\\
0 & b_2 & 0 \\
0 & 0 & c_3
\end{array}
\right), M^{(42)}_{D}=\left(
\begin{array}{ccc}
0 & a_2 & a_3\\
0 & b_2 & 0 \\
0 & 0 & c_3
\end{array}
\right), \\ \nonumber & M^{(43)}_{D}=\left( 
\begin{array}{ccc}
0 & a_2 & 0\\
0 & b_2 & b_3 \\
0 & 0 & c_3
\end{array}
\right), M^{(44)}_{D}=\left(
\begin{array}{ccc}
0 & a_2 & 0\\
b_1 & b_2 & 0 \\
0 & 0 & c_3
\end{array}
\right).
\label{eq:tautau_fv3}
\end{eqnarray}
\vspace{-0.7cm}
\begin{eqnarray} \label{eq:tautau_fv4}
M^{(3)}_S  ,&M^{(45)}_{D}= \left(
\begin{array}{ccc}
0 & a_2 & a_3\\
0 & 0 & b_3 \\
0 & c_2 & 0
\end{array}
\right), M^{(46)}_{D}=\left(
\begin{array}{ccc}
a_1 & 0 & a_3\\
0 & 0 & b_3 \\
0 & c_2 & 0
\end{array}
\right), \nonumber \\ &M^{(47)}_{D}=\left(
\begin{array}{ccc}
0 & 0 & a_3\\
b_1 & 0 & b_3 \\
0 & c_2 & 0
\end{array}
\right),M^{(48)}_{D}=\left(
\begin{array}{ccc}
0 & 0 & a_3\\
0 & b_2 & b_3 \\
0 & c_2 & 0
\end{array}
\right).
\end{eqnarray}

Note that in general the  entries of the 
Yukawa matrices $M_D$, $M_R$ and $M_S$ are complex (of the form 
$p e^{i \theta}$). However some of the phases can be absorbed by 
redefinition of  the leptonic fields. For the case 
when $M_R$ is diagonal, the number of un-absrobed phases is two --   
one each in $M_D$ and $M_S$ whereas for the  off-diagonal $M_R$  
only one phase remains in $M_S$. In this section we do 
not explicitly write the phases. However in section \ref{sec5} where we discuss specific
cases, the phases are explicitly included.

\section{Active neutrino mass matrix with one zero texture}
\label{sec4}

 The ($ 3\times3 $) light neutrino mass matrix being symmetric, there are 6 possible cases of one
zero textures with a vanishing lowest mass and these are studied in details in 
 Refs. \cite{Merle:2006du,Lashin:2011dn,Gautam:2015kya,Lavoura:2013ysa}. In the above section we observed
 that in context of MES model only viable textures of 
$m_\nu$ that we obtain are 
$ m_{e \tau} = 0$ and $ m_{\tau \tau} = 0$. 
According to the recent studies \cite{Gautam:2015kya,Lavoura:2013ysa,Harigaya:2012bw},
both these textures are ruled out for normal hierarchy when the lowest mass 
$m_1$ is zero but they can be allowed for the inverted hierarchy even when then lowest mass $m_3$ is zero
\footnote{We also observed that both these textures are disallowed for NH with the most recent data.}.
This kind of mass pattern can be obtained completely from group 
theoretical point of view  if one assumes that Majorana neutrino
mass matrix displays flavor antisymmetry under some discrete subgroup of SU(3)
as discussed in \cite{Joshipura:2015zla,Joshipura:2016hvn}.
In this section we re-analyse the textures $m_{e \tau}=0$ and $m_{\tau \tau}=0$ for the inverted hierarchical mass spectrum
assuming $m_3=0$ in the light of 
recent neutrino oscillation data as given in Table(\ref{parameters}). 
In our analysis we find that correlations among various oscillation 
parameters become highly constrained as compared to the earlier
studies. This is due to the recent constraints on the 3$ \sigma $ ranges of the mass squared 
differences and $\theta_{13}$  as compared to earlier
results in \cite{Lashin:2011dn,Gautam:2015kya,Lavoura:2013ysa} \footnote{
The latest constraint on $ |\Delta m^{2}_{31}| $ comes from
T2K and NO$\nu$A  including both  appearance and disappearance modes 
\cite{Abe:2015awa, Salzgeber:2015gua, Adamson:2016tbq, Adamson:2016xxw}. 
Whereas reanalysis of KamLAND data shows decrease  in the value of $ \Delta m^{2}_{21} $ and $ \sin^{2} \theta_{12} $ as discussed in \cite{Capozzi:2016rtj}.
}. 

In three neutrino paradigm, low energy Majorana neutrino mass matrix can be diagonalized as,
\begin{equation}
m^{3\times3}_{\nu}=U^{\prime}diag(m_{1},m_{2},m_{3} )U^{\prime T}.
\label{low_mnu}
\end{equation}
Here, $U^{\prime}=U.P$ ($ P=diag(1,e^{i \alpha}, e^{i (\beta + \delta_{13}) }) $) is a lepton mixing matrix in the basis where $M_l$ is diagonal.
The Pontecorvo-Maki-Nakagawa-Sakata (PMNS) matrix $U$ has 3 mixing angles and a CP violation phase $ \delta_{13} $.    
%
%
%
%
%
%

The elements of neutrino mass matrix can be calculated from 
Eq.(\ref{low_mnu}) are,
\begin{equation}\label{1zerotex}
(m^{3\times3}_{\nu})_{ab} = m_1 U_{a1}U_{b1}+m_2  U_{a2}U_{b2}e^{2 i \alpha}+m_3  U_{a3}U_{b3}e^{2 i( \beta+\delta_{13})},
\end{equation}
where, $ a,b=e,\mu~ and ~\tau $ and $ m_i (i = 1, 2,3) $ are given in Table(\ref{mass}). We express elements of $m_\nu$ as $m_{ab}$ in the text.\\

Imposing the condition of zero texture for IH 
with $m_3=0$ in the above equation we get, 
\begin{equation}\label{1zerotex_con}
 m_1 U_{a1}U_{b1}+m_2  U_{a2}U_{b2}e^{2 i \alpha} = 0,
\end{equation}
which can be simplified to obtain the mass ratio
\begin{equation}\label{mass_ratio1}
\frac{m_{1}}{m_{2}} e^{-2i\alpha} = - \frac{U_{a2} U_{b 2}}{U_{a1} U_{b 1}}.
\end{equation}
Let, $ q = \frac{m_1}{m_2} e^{ - 2 i \alpha } $ we get \\
\begin{align} 
\alpha & = - \frac{1}{2} Arg(q), \\ 
|q| &=\frac{m_1}{m_2} = \left|  - \frac{U_{a2} U_{b 2}}{U_{a1} U_{b 1}}  \right|.  \label{q1}
\end{align}
Let us define the ratio of the two mass squared differences as,
\begin{align}
R_{\nu }& = \frac{\Delta m^{2}_{21}}{|\Delta m^{2}_{31} |} = \frac{1 - |q|^{2}}{ |q|^{2}}.
 \label{q2}
\end{align}
The $R_\nu$ defined above can be calculated either using the current neutrino mass
squared differences as given in Table(\ref{parameters}) or by calculating $|q|$. If the value
of $R_\nu$ calculated using $|q|$ falls in the allowed $3 \sigma$ range of $R_\nu$ from the current data,
then we say the texture under consideration is allowed by the current data. As given in Table(\ref{parameters}) we vary the Dirac CP
phase $\delta_{13}$ from 0$^\circ$ $<$ $\delta_{13}$ $<$ 360$^\circ$ while the relevant Majorana phase $\alpha$
in the range 0$^\circ$ $<$ $\alpha$ $<$ 180$^\circ$ and find the correlations among different parameters, specially
the predictions for $\alpha$ and $\delta_{13}$.

We also study the effective Majorana neutrino mass, $ m_{ee} $, governing neutrinoless double beta decay ($0\nu \beta\beta$)  for these allowed textures. 
 In three flavor paradigm this can be written as,
\begin{align}
\label{effectivemm}
m_{ee} &= \nonumber |\Sigma U_{ei}^2 m_i| \nonumber \\ 
&= | m_1 c_{12}^2c_{13}^2 +m_2 e^{2i\alpha}c_{13}^2s_{12}^2
+ m_3 e^{2i\beta}s_{13}^2| .
\end{align}
where $ c_{ij}(s_{ij}) = \cos\theta_{ij}(\sin\theta_{ij}),~ (i< j,~ i,j=1,2,3 $). From the above equation we understand that $m_{ee}$ depends on the Majorana phases but not on the Dirac phase.
Various experiments  such as
CUORE \cite{Gorla:2012gd}, GERDA \cite{Wilkerson:2012ga},
SuperNEMO \cite{Barabash:2012gc}, KamLAND-ZEN \cite{Gando:2012zm} and 
EXO \cite{Auger:2012ar} are 
looking for signatures for neutrinoless double beta decay ($0\nu\beta\beta$). 
The current experiments provide bounds on the effective Majorana mass $m_{ee}$
from the non-observation of $0\nu\beta\beta$. 
For instance, the combined results from KamLAND-ZEN and
EXO-200 \cite{Gando:2012zm} give the upper bound on the effective Majorana 
neutrino mass as
$m_{ee}<$ (0.12 - 0.25) eV where the range 
signifies the uncertainty in the nuclear matrix elements.
The future experiments can improve this limit by one order of magnitude. 
Below we discuss the various correlations that we obtain for the allowed 
textures. 

\subsection{Case I: $ m_{e \tau} = 0 $}
\begin{figure}[!h]
\centering
\includegraphics[width=4.7cm, height=3.5cm]{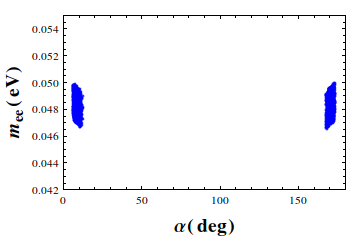}
\includegraphics[width=4.7cm, height=3.5cm]{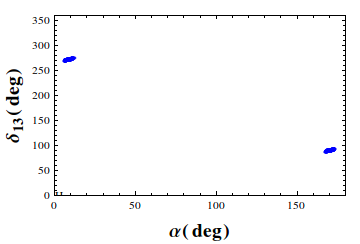}
\includegraphics[width=4.7cm, height=3.5cm]{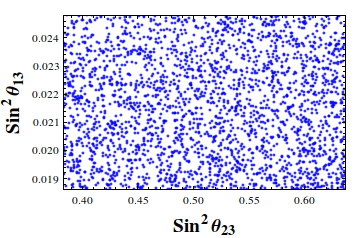}
\caption{Correlation plots of $m_{e \tau}$ = 0 for IH with vanishing $m_3$ in 3 neutrino paradigm.}
\label{metauih}
\end{figure}
The Majorana mass matrix element $ m_{e \tau} $ in 3-flavor case can be written as, 
\begin{equation}\label{1zerotex_etau}
m_{e \tau} = m_1 U_{e 1}U_{\tau 1}+m_2  U_{e 2}U_{\tau 2}e^{2 i \alpha}+m_3  U_{e 3}U_{\tau 3}e^{2 i( \beta+\delta_{13})}.
\end{equation}
Imposing the condition of zero texture with vanishing lowest mass ($ m_3 = 0$) for IH, we get,
\begin{align}
|m_1 U_{e 1}U_{\tau 1}+m_2  U_{e 2}U_{\tau 2}e^{2 i \alpha}| = 0, \\
|m_1 c_{12}c_{13} (s_{12}s_{23} - c_{12}c_{23}s_{13}e^{i\delta})+m_2 s_{12}c_{13} (-c_{12}s_{23} - 
s_{12}c_{23}s_{13}e^{i\delta})^2 e^{2i\alpha}| = 0.
\end{align}
From the above equation we obtain the mass ratio as below
\begin{equation}
\frac{m_2}{m_1} \approx 1- \frac{s_{13} \cos\delta_{13}}{\tan\theta_{23} s_{12}c_{12}}+\mathcal{O}(s_{13}^2).
\label{massratiotau}
\end{equation}

The mass ratio $\frac{m_2}{m_1}$ should be greater than 1. For this to happen $\cos \delta_{13}$ should be negative.
We find that 
due to the interplay
of the terms $\mathcal{O}(s_{13})$ and
$\mathcal{O}(s_{13}^2)$ the phase
$\delta_{13}$ is restricted 
to the range $[85^\circ-95^\circ]$ and $[265^\circ-275^\circ]$. 
The effective mass, $m_{ee}$ as function of  Majorana phase 
$\alpha$ is constrained due to very
small allowed range of $\alpha$ (5$^\circ < \alpha <$ 10$^\circ$, 170 $^\circ < \alpha < $ 175$^\circ$) as shown in  Eq(\ref{effectivemm}). 
The allowed range  of $m_{ee}$ for this texture is 0.046 eV $< m_{ee}<$ 0.05 eV and which can be probed in future experiments. 
Also, this texture predicts Dirac CP phase $ \sim 270^\circ$
which is in agreement with the indications from the 
current ongoing oscillation experiments like T2K and NO$\nu$A. 
There is however no constrain on the values of the neutrino mixing angles $\theta_{13}$ and $\theta_{23}$ seen in 
right panel of Fig.\ref{metauih} for this texture.

\subsection{Case II: $m_{\tau \tau} = 0 $}
The Majorana mass matrix element $ m_{\tau \tau} $ in 3-flavor case can
be written as, 

\begin{equation}\label{1zerotex_etau}
m_{\tau \tau} = m_1 U^{2}_{\tau 1}+m_2  U^{2}_{\tau 2}e^{2 i \alpha}+m_3  U^{2}_{\tau 3}e^{2 i( \beta+\delta_{13})}.
\end{equation}
Imposing the condition of texture zero with vanishing lowest mass($ m_3 = 0$)
for IH, we get,
\begin{align}
|m_1 U^{2}_{\tau 1}+m_2  U^{2}_{\tau 2}e^{2 i \alpha} | = 0, \\
|m_1 (s_{12}s_{23} - c_{12}c_{23}s_{13}e^{i\delta})^2  
 + m_2 (-c_{12}s_{23} - s_{12}c_{23}s_{13}e^{i\delta})^2 e^{2i\alpha}| = 0.
\end{align}
The mass ratio from the above equation can be written as
\begin{equation}
\frac{m_2}{m_1} \approx \frac{s_{12}^2}{c_{12}^2}\left[ 1-\frac{2 \cot\theta_{23}s_{13} 
\cos\delta_{13}}{c_{12} s_{12}}\right]+\mathcal{O}(s_{13}^2).
\label{massratiotau}
\end{equation}
\begin{figure}[!h]
\begin{center}
\includegraphics[width=6.5cm, height=4.5cm]{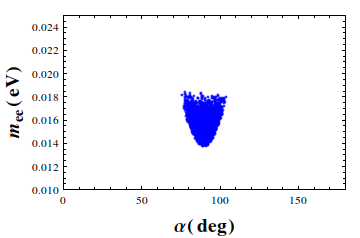}
\includegraphics[width=6.5cm, height=4.5cm]{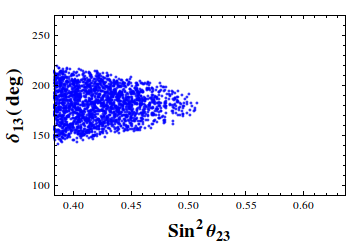}
\caption{Correlation plots of $m_{\tau \tau}$ = 0 for IH with vanishing $m_3$ in 3 neutrino paradigm.}
\label{figtautauih}
\end{center}
\end{figure}
Since this mass ratio $\frac{m_2}{m_1}$ is always greater than 1 
from oscillation data, we find that $\cos \delta_{13}$ should be negative for this texture as well. As can be seen from Fig \ref{figtautauih} that $\delta_{13}$ is constrained 
in the range  $140^{\circ}<\delta_{13}< 220^{\circ}$. We observe that, due to the more  constrained values  of mass squared differences  and $\theta_{13}$
from present data, as considered in our analysis, the atmospheric mixing angle $\theta_{23}$ is restricted to be below maximal. In the earlier analysis \cite{Lashin:2011dn,Gautam:2015kya,Lavoura:2013ysa}  there was no preferred octant of $\theta_{23}$.
The values of $\theta_{23}>$ 45$^\circ$ are disallowed for this texture as can
be seen in Fig \ref{figtautauih}. The effective mass, $m_{ee}$, being function of unknown Majorana phase  $\alpha$ as seen in Eq(\ref{effectivemm})
is constrained due to very  small allowed range of $\alpha$ (80$^\circ < \alpha <$110$^\circ$).
 The allowed range of $m_{ee}$ for this texture is 0.014 eV $< m_{ee}<$ 0.018 eV which is smaller compared to the case $m_{e\tau}$=0 where a vanishing element is off-diagonal.
The allowed values of the effective mass $m_{ee}$ for diagonal texture
$m_{\tau \tau}$ are on the lower side having no overlap with non diagonal
texture zero $m_{e\tau}$. Thus, $m_{ee}$ can be used to distinguish between diagonal and off-diagonal one texture zero classes with a vanishing neutrino mass. Note that allowed ranges of $\delta_{13}$ and $m_{ee}$ are more constrained in our
analysis as compared to references \cite{Lashin:2011dn,Gautam:2015kya} again due to 
the recent improved constraints on the
mass squared differences and $\theta_{13}$ at 3$\sigma$.

\section{Comparison of low and high energy neutrino mass matrix elements }
\label{sec5}

In this section we obtain the light neutrino  neutrino mass matrix ($m_\nu$) (Eq.\ref{mnu}), sterile mixing matrix ($m_s$) (Eq.\ref{mss}) and the active sterile mixing matrix (R) (Eq.\ref{rr}) using the different forms of $M_D$, $M_S$ and $M_R$ given in section (III) of the MES model. Since in the MES model both the active neutrino mass matrix $m_\nu$
and the active sterile mixing matrix $R$ depends on the parameters of
$M_S$, $M_D$ and $M_R$, this can induce additional correlations between active and sterile 
sector. Similarly, the mass of the sterile neutrino $m_s$ depends on 
$M_S$ and $M_R$. Hence expressing the various variables in terms 
of the parameters of these matrices one can get some interrelations.
 
For an illustration we will discuss three specific cases.
In case I and II we discuss $ m_{e \tau}=0 $ assuming diagonal structure of $ M_{R} $ and in
the case III we talk about $ m_{\tau \tau}=0 $ 
by considering the off diagonal form of $ M_{R} $. Note that here we consider the complex phases in our calculation. 
We compare high energy mass matrix with low energy
mass matrix after the decoupling of the eV sterile neutrino as discussed 
in section II. 

\begin{itemize}
\item Case I : 
Considering the forms of $ M^{(1)}_{S} $, $ M^{(1)}_{D} $ and diagonal $ M_{R} $ from Eq.(\ref{etau_1}),
\begin{equation}
 M^{(1)}_{S} = (0,s_2,s_3 e^{i \rho_2}),  M^{(1)}_{D}=\left( \begin{array}{ccc}
0 & 0 & a_{3}\\
b_{1} & 0 & b_{3} e^{i \rho_1} \\
c_{1} & 0 & 0
\end{array} \right), M_{R}=diag(r_1,r_2,r_3)   
\end{equation} 
 and using them in Eqs(\ref{mnu}, \ref{mss} and \ref{rr}) we get the low energy neutrino mass matrix, the sterile mass
 and the active sterile mixing matrix as,
\begin{align}
m^{3\times3}_\nu &=\left( \begin{array}{ccc}
- \frac{a_3^2 s_2^2}{(r_3 s_2^2 + r_2 s_3^2 e^{2 i \rho_2})} & -\frac{a_3 b_3 e^{i \rho_1}  s^{2}_2 }{(r_3 s_2^2 + r_2 s_3^2 e^{2 i \rho_2})} & 0 \\
. &  -\frac{b^{2}_1}{r_1}-\frac{b_3^2 s_2^2 e^{2 i \rho_1}}{(r_3 s_2^2 + r_2 s_3^2 e^{2 i \rho_2})}  & -\frac{b_1 c_1}{r_1}\\
. & . &-\frac{ c^{2}_1}{r_1}
\end{array} \right),
\label{mnu_result_1} 
\\
m_s & =-\left(  \frac{s^{2}_2}{r_2} + \frac{s^{2}_3 e^{2 i \rho_2}}{r_3}\right) ~,~~
R = \left( \begin{array}{c}  
\frac{a_3 r_2 s_3 e^{i \rho_2}}{(r_3 s^{2}_2 + r_2 s^{2}_{3} e^{2 i \rho_2})} \\
\frac{b_3 r_2 s_3 e^{i (\rho_1 + \rho_2})}{(r_3 s^{2}_2 + r_2 s^{2}_{3} e^{2 i \rho_2})}\\
0
\end{array} \right)  = \left(
\begin{array}{c}
V_{e4} \\ V_{\mu 4} \\0
\end{array}
\right).
\label{mnu_result} 
\end{align}
From Eq.(\ref{mnu_result_1}) and (\ref{mnu_result}) it can be seen that
\begin{align} 
\frac{m_{\mu \tau} }{m_{\tau \tau}} & = \frac{b_{1}}{c_{1}}, \nonumber ~~~
\frac{V_{e4}}{V_{\mu 4}}  = \frac{a_{3}}{b_{3}} e^{- i \rho_1} = \frac{m_{ee}}{m_{e \mu}} \label{sum_1}  \\ 
\end{align}
Here $m_{ab}, a,b=e,\mu,\tau$ are the low energy neutrino mass matrix elements. 
The eigen values of $ m^{3\times3}_\nu $ will give the masses of the three active neutrinos. 
Note that, only allowed hierarchy in our case is IH and hence $ m_3 = 0 $ and 
$ m_s=m_4=\sqrt{\Delta m^{2}_{43}} $. 
From Eq.(\ref{sum_1}) we get,
\begin{align} \label{sum_etau}
\left| \frac{V_{e4}}{V_{\mu 4}} \right| &  =\left| \frac{m_{ee}}{m_{e \mu}}  \right|.
\end{align}
%
 We  find that the lhs of Eq.(\ref{sum_etau}) lies in the range (0.63 -- 3.06) whereas rhs lies in (3.9 - 5.9) in their 3$ \sigma $ range. This shows that there is no overlapping between lhs and rhs of Eq.(\ref{sum_etau}) and hence disallowed from current neutrino oscillation data.
We observe that out of 12 forms of $ M^{(i)}_{D}, (i=1,2,...12) $ as given in Eq.(\ref{etau_1} -\ref{etau_3}), 6 of them ($ M^{(2)}_{D} $, $ M^{(4)}_{D} $,
$ M^{(6)}_{D} $, $ M^{(8)}_{D} $, $ M^{(9)}_{D} $ and $ M^{(11)}_{D} $) do not lead to the correlation given in Eq.(\ref{sum_etau}) and these $ M^{(i)}_{D} $'s are not ruled out. Hence a detail analysis of one of these  $ M^{(i)}_{D} $'s is discussed below in Case II.
\item Case II : 
Considering the form of $ M^{(1)}_{S} $, $ M^{(2)}_{D} $ and diagonal $ M_{R} $ given in Eq.(\ref{etau_1}),
\begin{equation}
 M^{(1)}_{S} = (0,s_2,s_3 e^{i \rho_2}),  M^{(2)}_{D}=\left( \begin{array}{ccc}
0 & a_{2} & 0\\
b_{1} & 0 & b_{3} e^{i \rho_1} \\
c_{1} & 0 & 0
\end{array} \right), M_{R}=diag(r_1,r_2,r_3)   
\end{equation} 
 and using them in Eqs(\ref{mnu}, \ref{mss} and \ref{rr}) we get the texture $m_{e\tau} = 0$,
\begin{align}
m^{3\times3}_\nu &=\left( \begin{array}{ccc}
- \frac{a_2^2 s_3^2 e^{2 i \rho_2}}{(r_3 s_2^2 + r_2 s_3^2 e^{2 i \rho_2})} & \frac{a_2 b_3 s_2 s_3 e^{i (\rho_1+ \rho_2)}}{(r_3 s_2^2 + r_2 s_3^2 e^{2 i \rho_2})} & 0 \\
. &  -\frac{b^{2}_1}{r_1}-\frac{b_3^2 s_2^2 e^{2 i \rho_1}}{(r_3 s_2^2 + r_2 s_3^2 e^{2 i \rho_2})}  & -\frac{b_1 c_1}{r_1}\\
. & . &-\frac{ c^{2}_1}{r_1}
\end{array} \right).
\label{mnu_result_2} 
\end{align}
The sterile mass and active sterile mixing becomes
\begin{align}
m_s & =-\left(  \frac{s^{2}_2}{r_2} + \frac{s^{2}_3 e^{2 i \rho_2}}{r_3}\right) ~,~~
R = \left( \begin{array}{c}  
\frac{a_2 r_3 s_2}{(r_3 s^{2}_2 + r_2 s^{2}_{3}e^{2 i \rho_2})} \\
\frac{b_3 r_2 s_3 e^{ i(\rho_1 + \rho_2)}}{(r_3 s^{2}_2 + r_2 s^{2}_{3}e^{2 i \rho_2})}\\
0
\end{array} \right)  = \left(
\begin{array}{c}
V_{e4} \\ V_{\mu 4} \\0
\end{array}
\right).
\end{align}
It can be seen from the above equations that
\begin{align}\label{eq:rel_1}
\frac{m_{\mu \tau} }{m_{\tau \tau}} & = \frac{b_{1}}{c_{1}},~~~\frac{m_{e e}}{m_{e \mu}} = - \frac{a_2  s_3}{b_3 s_2} e^{ i (\rho_2 - \rho_1)}
\end{align}
From Eq.(\ref{mnu_result_2}) we get the following relation between the light neutrino mass matrix elements,
\begin{align}
m_{\mu \mu} & = \frac{b_1}{c_1}m_{\mu \tau} - \frac{b_3 s_2}{a_2 s_3}e^{ i (\rho_1 - \rho_2)}m_{e \mu}
= \frac{m^{2}_{e \mu}}{m_{e e}} + \frac{m^{2}_{\mu \tau}}{m_{\tau \tau}} \nonumber
\end{align}
which implies,
\begin{align}\label{eq:mee_md2}
m_{ee} = \frac{m^{2}_{e \mu}m_{\tau \tau} }{m_{\mu \mu}m_{\tau \tau} - m^{2}_{\mu \tau}}.
\end{align}
To obtain Eq.(\ref{eq:mee_md2}) we have  used the correlations of 
Eq.(\ref{eq:rel_1}). 
Now to test the viability of these structures of $M_D$, $M_R$ and $M_S$, 
we look for the parameter space in which both the conditions $m_{e\tau}=0$ and Eq.(\ref{eq:mee_md2}) are satisfied simultaneously. 
In the upper panels of Fig. \ref{fig:metau_st_ih}, we have plotted the correlations obtained between different low energy parameters in this scenario.  
Comparing these correlations with Fig. \ref{metauih} 
(which corresponds to only $m_{e\tau}=0$), we find that the 
MES model
disfavours a large area in the $\sin^2 \theta_{23} - \sin^2 \theta_{13}$ plane 
and allows 
$\theta_{23}$  values  in the lower octant :  $0.383 < \sin^2 \theta_{23} < 0.42$   
whereas the admissible values of  
$\theta_{13}$  ( $ 0.021 < \sin^2\theta_{13} < 0.0248$) are 
 near the higher side of it's allowed range.
However the values of $\alpha$ and $m_{ee}$ which are predicted by the
two cases are similar. The prediction of the texture with  $m_{e\tau}=0$ 
is $ 6^\circ < \alpha < 13^\circ $ and $167^\circ < \alpha < 174^\circ$ 
while the MES model predicts a slightly constrained range 
$11.7^\circ < \alpha < 13^\circ$ and $167^\circ < \alpha < 168.1^\circ$.
In this case we also obtain another correlation for sterile neutrino mass from this model of the form,
\begin{equation}\label{ms_metau}
m_{s} =\left| - \frac{m_{e \mu}}{V_{e4}V_{\mu 4}} \right|.
\end{equation}
\begin{figure}[!h]
\begin{center}
\includegraphics[width=6.5cm, height=4.5cm]{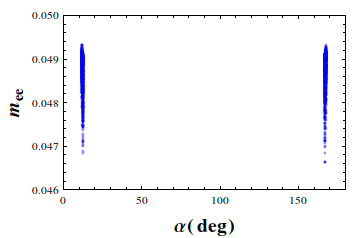}
\includegraphics[width=6.5cm, height=4.5cm]{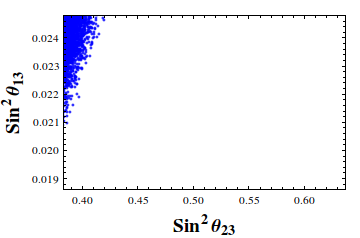}
\includegraphics[width=6.5cm, height=4.5cm]{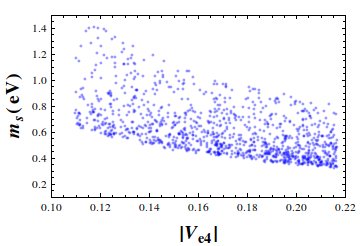}
\includegraphics[width=6.5cm, height=4.5cm]{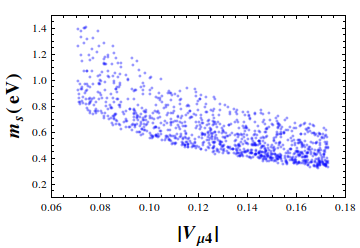}
\caption{ Correlation plots for case II.}
\label{fig:metau_st_ih}
\end{center}
\end{figure}
 In the lower panels of Fig.\ref{fig:metau_st_ih}, we have plotted the prediction of $m_s$ as given by Eq.(\ref{ms_metau}) by varying 
$V_{e4}$ and $V_{\mu 4}$ within their allowed range as given in table(\ref{parameters}). This is obtained when both the conditions i.e., $m_{e \tau}=0$ and Eq.(\ref{eq:mee_md2}) is satisfied simultaneously. From the figures we see that the prediction of $m_s$ by this model is consistent with data coming from the SBL experiments.
\item Case III : 
Considering the cases for  the off-diagonal forms of  $M_R$ given in
Eqs.(\ref{eq:e_tau_1_fv1}-\ref{eq:tautau_fv4}),
we find that out of the 36 $ M^{(i)}_{D}, (i=13, 14,..., 48) $,  19 cases
lead to  exactly  the same correlation depicted by Eq.(\ref{sum_etau}). 
This is not allowed from current oscillation data  as discussed earlier. 
Among the remaining 17 cases 11 
$M^{(i)}_{D}$ (for $i=17, 19, 22, 23, 30, 32, 34, 42, 43, 44 $ and $ 45$) 
lead  to a correlation of the form,
\begin{equation}\label{eq:sum_etau_2}
\left| \frac{V_{e4}}{V_{\mu 4}} \right| = \left| \frac{m_{e \tau}}{m_{\mu \tau}} \right|.
\end{equation}
This  is also not satisfied by current neutrino oscillation data
as  the rhs of Eq.(\ref{eq:sum_etau_2}) lies in the range (3.8 -- 4.7) 
showing no overlapping with lhs. 
The remaining six forms of $M^{(i)}_{D}$ are  $ M^{(18)}_{D} $, $ M^{(21)}_{D} $, $ M^{(29)}_{D} $, $ M^{(33)}_{D} $, $ M^{(41)}_{D} $ and $ M^{(46)}_{D} $. All these forms of $ M_{D} $ and the corresponding forms of $ M_R $ and $ M_S $ lead to neutrino mass matrix with $ m_{\tau \tau} = 0 $. We found that, all these $ M_{D} $'s lead a correlation of  the form,
\begin{equation} \label{eq:sum_tautau}
\left| \frac{V_{e4}}{V_{\mu 4}} \right|  =\left| \frac{m_{e \mu}}{m_{\mu \mu}}  \right|
\end{equation}
which is satisfied by current oscillation data. 
The rhs of Eq.(\ref{eq:sum_etau_2}) lies in the range (1.8 -- 2.3) which shows
complete overlap with lhs (0.63- 3.06). 
For illustration, we consider  $ M^{(18)}_{D} $ with corresponding $ M_{R} $ and $ M^{(1)}_{S} $  and using them in Eqs(\ref{mnu}, \ref{mss} and \ref{rr}) we get ,
\begin{align}
m^{3\times3}_\nu &=\left( \begin{array}{ccc}
\frac{a_1 s_2(a_1 s_2 r_1+2 a_3 s_3 r_2 e^{i \rho_2})}{r_2^2 s_3^2 e^{2 i \rho_2}} & \frac{b_1 s_2(a_1 s_2 r_1 +a_3 s_3 r_2 e^{ i \rho_2})}{r_2^2 s_3^2 e^{2 i \rho_2}} & -\frac{a_1 c_2}{r_2} \\
. &  \frac{b^{2}_1 s_2^2 r_1}{r_2^2 s_3^2} e^{- 2 i \rho_2}  & -\frac{b_1 c_2}{r_2}\\
. & . &0
\end{array} \right),
\label{mnu_result_3} 
\\
m_s & =- \frac{s^{2}_3}{r_1} e^{2 i \rho_2}~,~~
R = \left( \begin{array}{c}  
\frac{a_1  s_2 r_1 +  a_3 s_3 r_2 e^{ i \rho_2}}{r_2 s^{2}_3 e^{2 i \rho_2}} \\
\frac{b_1 s_2  r_1}{r_2 s^{2}_3 e^{2 i \rho_2}}\\
0
\end{array} \right)  = \left(
\begin{array}{c}
V_{e4} \\ V_{\mu 4} \\0
\end{array}
\right).
\end{align}
From the above matrices we find the following correlation,
\begin{align}
m_{s}  =\left| - \frac {m_{e \mu}}{V_{e4}V_{\mu 4}} \right| \label{eq:ms_tautau}.
\end{align}
also the correlation mentioned by Eq.(\ref{eq:sum_tautau}). We find  that  both the  equations (\ref{eq:sum_tautau} and \ref{eq:ms_tautau})  are consistent with the current oscillation data. The simultaneous validity of equations (\ref{eq:sum_tautau} and \ref{eq:ms_tautau}) lead to light sterile neutrino mass in the range $ 1.4~ eV < m_s < 3.5~ eV $ which is marginally allowed 
by global analysis as seen from Fig.(\ref{fig:mtautau_st_ih}). However, individual experiments (MINOS, IceCube, Daya Bay) still allow higher value of sterile neutrino mass 
\cite{An:2016luf,TheIceCube:2016oqi, MINOS:2016viw, Adamson:2016jku}. 
\end{itemize}
\begin{figure}[!h]
\begin{center}
\includegraphics[width=6.5cm, height=4.5cm]{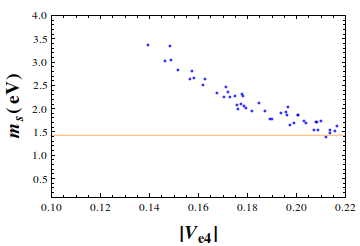}
\includegraphics[width=6.5cm, height=4.5cm]{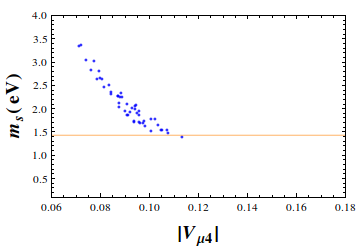}
\caption{Sterile neutrino mass from Eqs.(\ref{eq:ms_tautau}) for $ m_{\tau \tau} = 0 $.
The yellow line is the current upper bound on $ m_s $ as given by global analysis of 3+1 neutrino oscillation data. }
\label{fig:mtautau_st_ih}
\end{center}
\end{figure}
In Table(\ref{table:allowed_result_etau}) and Table(\ref{table:allowed_result_tautau}) we 
summarize the allowed cases that we obtained in our study for texture $ m_{e \tau} = 0 $
and $ m_{\tau \tau} = 0 $ respectively.
\begin{table}[!h]
\centering
\begin{tabular}{|c|c|c|c|c|}
\hline
Case & $ M_S $ & $ M_D $ & $ M_R $ & Correlations \\ 
\hline
I & $(0,s_2,s_3 )$  & $\left( \begin{array}{ccc}
0 & a_{2} & 0\\
b_{1} & 0 & b_{3} \\
c_{1} & 0 & 0
\end{array} \right)$ & diag($r_1, r_2, r_3$) &  $m_{ee} = \frac{m^{2}_{e \mu}m_{\tau \tau} }{m_{\mu \mu}m_{\tau \tau} - m^{2}_{\mu \tau}}$\\
& & & & $ m_{s} =\left| - \frac{m_{e \mu}}{V_{e4}V_{\mu 4}} \right|$ \\
\hline
II & $(0,s_2,s_3)$  & $\left( \begin{array}{ccc}
0 & 0 & a_{3}\\
b_{1} & b_2 & 0 \\
c_{1} & 0 & 0
\end{array} \right)$ & diag($r_1, r_2, r_3$) &  Same as Case I \\
\hline
III & $(s_1,0,s_3)$  & $\left( \begin{array}{ccc}
0 & 0 & a_{3}\\
0 & b_2 & b_3 \\
0 & c_2 & 0
\end{array} \right)$ & diag($r_1, r_2, r_3$) &  Same as Case I \\
\hline
IV & $(s_1,0,s_3)$  & $\left( \begin{array}{ccc}
a_1 & 0 & 0\\
b_2 & b_2 & 0 \\
0 & c_2 & 0
\end{array} \right)$ & diag($r_1, r_2, r_3$) &  Same as Case I \\
\hline
V & $(s_1,s_2,0)$  & $\left( \begin{array}{ccc}
0 & a_2 & 0\\
0 & b_2 & b_3 \\
0 & 0 & c_3
\end{array} \right)$ & diag($r_1, r_2, r_3$) &  Same as Case I \\
\hline
VI & $(s_1,s_2,0)$  & $\left( \begin{array}{ccc}
a_1 & 0 & 0\\
b_1 & 0 & b_3 \\
0 & 0 & c_3
\end{array} \right)$ & diag($r_1, r_2, r_3$) &  Same as Case I \\
\hline
\end{tabular}
\begin{center}
\caption{The various forms of $ M_D $, $ M_R $ and $ M_S $ which leads to a phenomenologically allowed $ m_{e \tau} = 0 $. }
\label{table:allowed_result_etau}
\end{center}
\end{table}
\begin{table}[!h]
\centering
\begin{tabular}{|c|c|c|c|c|}
\hline
Case & $ M_S $ & $ M_D $ & $ M_R $ & Correlations \\ 
\hline
I & $(0,s_2,s_3)$  & $\left( \begin{array}{ccc}
a_1 & 0 & a_3\\
b_{1} & 0 & 0 \\
0 & c_2 & 0
\end{array} \right)$ & $\left( \begin{array}{ccc}
0 & r_2 & 0\\
r_2 & 0 & 0 \\
0 & 0 & r_1
\end{array} \right)$ &  $m_{s} =\left| - \frac{m_{e \mu}}{V_{e4}V_{\mu 4}} \right|$ \\
\hline
II & $(s_1,0,s_3)$  & $\left( \begin{array}{ccc}
a_1 & a_2 & 0\\
0 & b_2 & 0 \\
c_{1} & 0 & 0
\end{array} \right)$ & $\left( \begin{array}{ccc}
0 & r_2 & 0\\
r_2 & 0 & 0 \\
0 & 0 & r_1
\end{array} \right)$ &  Same as Case I \\
\hline
III & $(0,s_2,s_3)$  & $\left( \begin{array}{ccc}
a_1 & a_2 & 0\\
b_1 & 0 & 0 \\
0 & 0 & c_3
\end{array} \right)$ & $\left( \begin{array}{ccc}
0 & 0 & r_2\\
0 & r_1 & 0 \\
r_2 & 0 & 0
\end{array} \right)$ &  Same as Case I \\
\hline
IV & $(s_1,s_2,0)$  & $\left( \begin{array}{ccc}
0 & a_2 & a_3\\
0 & 0 & b_3 \\
c_1 & 0 & 0
\end{array} \right)$ & $\left( \begin{array}{ccc}
0 & 0 & r_2\\
0 & r_1 & 0 \\
r_2 & 0 & 0
\end{array} \right)$ &  Same as Case I \\
\hline
V & $(s_1,0, s_3)$  & $\left( \begin{array}{ccc}
a_1 & a_2 & 0\\
0 & b_2 & b_3 \\
0 & 0 & c_3
\end{array} \right)$ &  $\left( \begin{array}{ccc}
r_1 & 0 & 0\\
0 & 0 & r_2 \\
0 & r_2 & 0
\end{array} \right)$ &  Same as Case I \\
\hline
VI & $(s_1,s_2, 0)$  & $\left( \begin{array}{ccc}
a_1 & a_2 & 0\\
0 & b_2 & b_3 \\
0 & 0 & c_3
\end{array} \right)$ & $\left( \begin{array}{ccc}
r_1 & 0 & 0\\
0 & 0 & r_2 \\
0 & r_2 & 0
\end{array} \right)$ &  Same as Case I \\
\hline
\end{tabular}
\begin{center}
\caption{The various forms of $ M_D $, $ M_R $ and $ M_S $ which leads to a phenomenologically allowed $ m_{\tau\tau} = 0 $.}
\label{table:allowed_result_tautau}
\end{center}
\end{table}
%
\subsection{NLO correction for MES model} \label{sec:p51}

\begin{figure}[!h]
\begin{center}
\includegraphics[width=6.5cm, height=4.5cm]{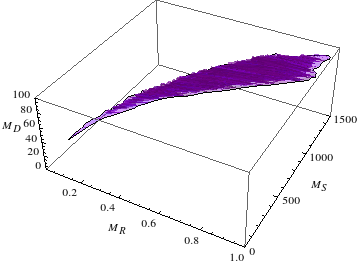}
\caption{This plot shows the allowed parameter spaces of $ M_D $ (GeV), $ M_R $ (in units of $10^{15}$ GeV) and $ M_S $ (GeV) which lead to NLO correction term$ \sim 10^{-5} $ eV or less.}
\label{fig:3dplot}
\end{center}
\end{figure}

In sec.(\ref{sec3}), the structures of various mass matrices are obtained using the leading order expression of $ m^{3\times3}_{\nu} $ as given by equation (\ref{mnu}) which give rise to texture zeros with exact cancellation. However, if $M_D/M_S\sim0.1$  NLO corrections can be important. In this section, we discuss  the effect of NLO correction terms for MES model corresponding to the allowed texture zeros.
The NLO correction term  can be calculated following the  standard algorithm given in \cite{Grimus:2000vj}. To calculate the NLO term, let us rewrite equation (\ref{mnus}) in the form,
\begin{equation}
M^{4\times4}_{\nu} = \left( \begin{array}{cc}
\mathcal{M}_L & \mathcal{M}^T_D\\
\mathcal{M}_D & \mathcal{M}_R  
\end{array} \right)
\end{equation}
where, 
\begin{equation}\label{eq:form_mdmrms}
 \mathcal{M}_L =  M_D M^{-1}_R  M^{T}_D, ~ \mathcal{M}_D = M_S (M^{-1}_R)^{T} M^{T}_D,~ \mathcal{M}_R =  M_S M^{-1}_R  M^{T}_S
\end{equation}
\begin{align} \label{eq:nlo_corr}
(m^{3\times3}_{\nu})_{NLO} & = \frac{1}{2}  \left[ \mathcal{M}^{T}_D \mathcal{M}^{-1}_R\mathcal{M}^{-1*}_R \mathcal{M}^{*}_D \mathcal{M}_L+ (last~ term)^{T} \right]  \nonumber \\
&-\frac{1}{2} \mathcal{M}^{T}_D \mathcal{M}^{-1}_R \left[  \mathcal{M}_D \mathcal{M}^{\dagger}_D \mathcal{M}^{-1*}_R + (last~ term)^{T}\right] \mathcal{M}^{-1}_R \mathcal{M}_D\nonumber \\
 & = \frac{1}{2} [ M_D M^{-1}_R  M^{T}_S~(M_S M^{-1}_R  M^{T}_S)^{-1} (M^*_S M^{-1*}_R  M^{\dagger}_S)^{-1}~M^*_S (M^{-1}_R)^{\dagger}  M^{\dagger}_D ~ M_D M^{-1}_R  M^{T}_D  \nonumber \\
&+(last~ term)^{T} ] \nonumber \\
& - \frac{1}{2} ~ M_D M^{-1}_R  M^{T}_S~(M_S M^{-1}_R  M^{T}_S)^{-1}[ M_S (M^{-1}_R)^{T}  M^{T}_D~~M^*_D (M^{-1}_R)^*  M^{\dagger}_S \nonumber \\
& (M_S M^{-1}_R  M^{T}_S)^{-1*}+ (last~ term)^{T}] (M_S M^{-1}_R  M^{T}_S)^{-1} ~M_D M^{-1}_R  M^{T}_S  
\end{align}
In the second line we use the form of $ \mathcal{M}_L,~ \mathcal{M}_D $ and $ \mathcal{M}_R $ as given by equation (\ref{eq:form_mdmrms}) to obtain the final form given by equation (\ref{eq:nlo_corr}). We see that the
contribution of the NLO terms of equation (\ref{eq:nlo_corr}) are proportional to $ M^{4}_D/M_R M^{2}_S $.
%
This implies that a term of the order $ M^{4}_D/M_R M^{2}_S $ will add to every term of $ m^{3\times3}_{\nu} $ as given by the equation (\ref{mnu_result_2}). To get the specific form of NLO correction term, in equation (\ref{eq:nlo_corr}), we use the specific forms of $ M_D $, $ M_R $ and $ M_S $ used for obtaining equation (\ref{mnu_result_2}). The NLO correction term we obtain for (1,3) element of equation (\ref{mnu_result_2}) is $\sim\frac{a_3 b_3 b_1 c_1 r^{2}_2 s^{2}_3}{2 r_1(r_3 s^{2}_2+r_2 s^{2}_3)^{2}} $,  which is of the  order of $ M^{4}_D/M_R M^{2}_S $, where $ a_3,~ b_3,~ b_1,~ c_1$ are elements of $ M_D $, $ r_1,~r_2 $ are elements of $ M_R $ and $ s_2,~s_3 $  are elements of $ M_S $. We see here that because of NLO corrections, we no longer  have exact cancellation leading to $ m_{e \tau} = 0$, unlike the leading order case. But, if we consider representative values of parameters say, $ M_D \sim 80$ GeV, $ M_R \sim 6\times10^{14}$ GeV and $ M_S \sim 1000$ GeV then  we find that $ m_{\nu}\sim0.011 $ eV,  $ m_{s} \sim 1.6 $ eV, R$\sim0.1  $ and NLO$ \sim 10^{-5} $ eV. In figure (\ref{fig:3dplot}) we show the allowed parameter spaces of $ M_D $, $ M_R $ and $ M_S $  which can lead to NLO correction term$ \sim 10^{-5} $ eV or less.\footnote{In our numerical analysis texture zero (say, $ m_{e \tau} = 0$) corresponds to $ m_{e \tau} = 10^{-5}~eV$.} Hence, there exist a parameter space where we can safely neglect NLO correction terms in our analysis compared to leading order terms and consider the texture zero even with the inclusion of the NLO term.\footnote{We notice that the set of $ M_D $, $ M_R $ and $ M_S $ which do not give NLO$ \sim 10^{-5} $ eV do not give the one zero textures.} Thus, all the model predictions corresponding to leading order terms remain unchanged. Note that similar  conclusions can also be obtained for the texture $ m_{\tau \tau} = 0$.
\section{Symmetry realization}
\label{sec6}

Singular one zero neutrino mass matrices can be realized using a discrete Abelian flavor symmetry within the context of MES mechanism. Earlier in \cite{Grimus:2004hf} authors studied the possibilities to enforce zero textures in arbitrary entries of the
fermion mass matrices by means of Abelian symmetries in the context of type - I seesaw mechanism. We adopt the same approach  to probe the zero textures of $m_{\nu}$ in the context of MES mechanism. We observe that one zero textures of $m_{\nu}$ with a vanishing mass can be realized by $Z_8 \times Z_2$ symmetry. To realize the texture structures we extend the SM particle composition by three right handed neutrinos ($\nu_{eR}$, $\nu_{\mu R}$, $\nu_{\tau R}$) as required in MES model and two more Higgs doublets ($\phi'$, $\phi''$) in addition to the SM one ($\phi$). 
Few $SU(2)_L$ scalar singlets ($\chi_i$, $i=1,2$) are required
to realize diagonal $M_R$ whereas two  singlets $\lambda_i$, $i=1,2$ helps in realizing one zero texture structure of $M_S$. 
Note that the model that we discuss here to get the zero texture structure is general, flexible and in no way unique.
The additional discrete group $ Z_2 $ is introduced to restrict some of the unwanted terms in the Lagrangian. 
For illustration, we present the detailed symmetry realization of our
two viable textures of $m_{\nu}$ $(m_{e \tau}, m_{\tau\tau} = 0)$. 
 The particle assignments for $(m_{e \tau}=0$ which is allowed by current data (case II)
 under the action of $Z_8 \times Z_2$ symmetry are given in Table (\ref{field_trans}).
\begin{table}[!h]
\centering
\begin{small}
\begin{tabular}{|c|c|c|c|c|c|c|c|}
\hline
Lepton  & ($ Z_{8}\times Z_{2} $) & RH Singlet &($ Z_{8}\times Z_{2} $) & $ \nu $ fields & ($ Z_{8}\times Z_{2} $) & Higgs & ($ Z_{8}\times Z_{2} $)\\ 
doublet & & & & & & doublet & \\
\hline
$ \bar{D}_{L_e} $ & ($ \omega^6 $, -1) & $ e_{R} $ & ($ \omega^2 $, -1) & $ \nu_{e R} $ & ($ \omega^5 $, 1) & $ \phi $&(1, 1)\\
$ \bar{D}_{L_\mu} $ & ($ \omega^3 $, 1) & $ \mu_{R} $ & ($ \omega^5 $, 1) & $ \nu_{\mu R} $ & ($ \omega^2 $, -1) & $ \phi^{\prime} $ &($ \omega^3 $, 1)\\
$ \bar{D}_{L_\tau} $ & ($ \omega^5$, 1) & $ \tau_{R} $ & (1, 1) & $ \nu_{\tau R} $ & (1, 1) & $ \phi^{\prime \prime} $ & ($ \omega^2 $, 1)\\
\hline
\end{tabular}
\begin{center}
\caption{Here, $\bar{D}_{L_l}$ 
denote $SU(2)_L$ doublets and $l_R$, $\nu_{l_R}$ ($l=e, \mu, \tau$)
are the right-handed (RH) $SU(2)_L$ singlet for charged lepton and neutrino fields respectively. 
Also, $ \phi, \phi^{\prime}$ and $\phi^{\prime\prime}$ are the Higgs doublets.}
\label{field_trans}
\end{center}
\end{small}
\end{table}

According to the charge assignments of the leptonic field given in Table (\ref{field_trans}) the bilinears $\bar{D}_{L_l}l_R$, $\bar{D}_{L_l}\nu_{l_R}$ and $\nu_{l_R}^T C^{-1}\nu_{l_R}$
relevant for $M_l$, $M_D$ and $M_R$ transform as,
\begin{center}
$\bar{D}_{L_l}l_R \sim \left(
\begin{array}{ccc}
1 & \omega^3 &\omega^6 \\
\omega^5 &1 &\omega^3 \\
\omega^7 & \omega^2 & \omega^5
\end{array}
\right)$,~
$\bar{D}_{L_l}\nu_{l_R} \sim \left(
\begin{array}{ccc}
\omega^3 & 1 & \omega^6 \\
1 & \omega^5 & \omega^3 \\
\omega^2 & \omega^7 & \omega^5
\end{array}
\right)$,~
$\nu_{l_R}\nu_{l^{\prime}_R} \sim \left( \begin{array}{ccc}
\omega^2 & \omega^7 &\omega^5 \\
\omega^7 & \omega^4 &\omega^2 \\
\omega^5 &\omega^2 & 1
\end{array}
\right)$,
\end{center} 
where $ \omega = e^{\pi i/4}, $ $\omega^8=1$ .
 We introduce three $SU(2)_L$ doublet Higgs ($ \phi, \phi^{\prime}$,$\phi^{\prime\prime}$). 
One of these Higgs doublet $\phi$, is invariant under $Z_8$ while the other two fields transforms as: 
 $\phi^{\prime} \rightarrow \omega^3 \phi^{\prime}$ ($\tilde{\phi^{\prime}}\rightarrow \omega^5 \tilde{\phi^{\prime}}$) and
 $\phi^{\prime\prime} \rightarrow \omega^2 \phi^{\prime \prime}$ ($\tilde{\phi^{\prime \prime}}\rightarrow \omega^6 \tilde{\phi^{\prime\prime}}$). 
 The ($Z_8 \times Z_2$) invariant Yukawa Lagrangian than becomes
\begin{align}
-\mathcal L_{Y}&= Y_{ee} \bar{D}_{L_e} e_R \phi + Y_{\mu \mu}\bar{D}_{L_\mu} \mu_R \phi +Y_{\tau \tau}\bar{D}_{L_\tau} \tau_R \phi^{\prime} + \\  \nonumber
 & Y_{e \mu} \bar{D}_{L_e} \nu_{\mu_R} \tilde{\phi}+Y_{\mu e} \bar{D}_{L_\mu} \nu_{e_R}\tilde{\phi}+Y_{\mu \tau} \bar{D}_{L_\mu} \nu_{\tau_R}\tilde{\phi^{\prime}}+Y_{\tau e} \bar{D}_{L_\tau} \nu_{e_R}\tilde{\phi^{\prime\prime}}+h.c..
\end{align}
here all $\tilde{\phi}=i \tau_2 \phi^*$. The Higgs fields acquires the vacuum expectation values 
$\langle\phi\rangle_o \neq 0$ and results in the $M_l$ and $M_D$ of the following form,
\be
M_l = \left(
\begin{array}{ccc}
m_e & 0 & 0 \\
0 & m_{\mu}& 0 \\
0 & 0 & m_{\tau}
\end{array}
\right),
M_D = \left(
\begin{array}{ccc}
0 & a_2 & 0 \\
b_1 & 0 & b_3 \\
c_1 & 0 & 0
\end{array}
\right).
\ee
Here $m_e = Y_{ee} \langle \phi \rangle_o$, $m_{\mu} = Y_{\mu \mu} \langle \phi \rangle_o$
$m_{\tau} = Y_{\tau \tau} \langle \phi^{\prime} \rangle_o$. The elements of $M_D$ are $a_2=Y_{e \mu}\langle \phi^* \rangle_o$,
$b_1=Y_{\mu e}\langle \phi^{ *} \rangle_o$, $b_3=Y_{\mu \tau}\langle \phi^{\prime  *} \rangle_o$ and
$c_1=Y_{\tau e}\langle \phi^{\prime \prime *} \rangle_o$. 
For the right-handed Majorana mass matrix ($ M_R $) and for the mass matrix $M_S$,  
we introduce few $SU(2)_L$ scalar singlets and their transformation under $Z_8 \times Z_2$
is given in the Table(\ref{sig_field_trans}).
\begin{table}[!h]
\centering
\begin{tabular}{|c|c|c|c|}
\hline
Scalar singlet & ($ Z_{8}\times Z_{2} $) & Scalar singlet & ($ Z_{8}\times Z_{2} $) \\ 
\hline
$ \chi_1 $ & ($ \omega^6 $, 1) & $ \lambda_{1} $ & (1, 1)  \\
$ \chi_2 $ & ($ \omega^4 $, 1) & $ \lambda_{2} $ &  ($ \omega^2 $, -1)  \\
\hline
\end{tabular}
\begin{center}
\caption{Here, scalar singlet $ \chi_1 $ and $ \chi_2 $ give $M_R$ whereas $ \lambda_{1} $ and $ \lambda_{2} $ give $M_S$.}
\label{sig_field_trans}
\end{center}
\end{table}
Thus the mass matrices $ M_R $ and $ M_S $ becomes,
\be
M_R = \left(
\begin{array}{ccc}
r_1 & 0 & 0 \\
0 & r_2 & 0 \\
0 & 0 & r_3
\end{array}
\right),~ M_S = \left(
\begin{array}{ccc}
 0 & s_2 & s_3
\end{array}
\right).
\ee
 We also give the transformation to the singlet field S as ($ \omega^6 $, -1)  under ($ Z_8 \times Z_2 $) which will prevent the term of the form $ \overline{S^{c}} S $ as demand by the MES model will still give the  correct form of $M_S$.

Using the minimal extended type I seesaw given in Eqn (\ref{mnu}) with the mass matrices
$M_D$, $M_R$ and $M_S$ as discussed above leads to effective neutrino mass matrix $m_{\nu}$ with a texture 
zero at (1,3) position.
\begin{table}[!h]
\centering
\begin{small}
\begin{tabular}{|c|c|c|c|c|c|c|c|}
\hline
Lepton  & ($ Z_{8}\times Z_{2} $) & RH Singlet &($ Z_{8}\times Z_{2} $) & $ \nu $ fields & ($ Z_{8}\times Z_{2} $) & Higgs & ($ Z_{8}\times Z_{2} $)\\ 
doublet & & & & & & doublet & \\
\hline
$ \bar{D}_{L_e} $ & (1, 1) & $ e_{R} $ & (1, 1) & $ \nu_{e R} $ & ($ \omega^3 $, 1) & $ \phi $&(1, 1)\\
$ \bar{D}_{L_\mu} $ & ($ \omega^5 $, -1) & $ \mu_{R} $ & ($ \omega^3 $, -1) & $ \nu_{\mu R} $ & ($ \omega^5 $, 1) & $ \phi^{\prime} $ &($ \omega^3 $, 1)\\
$ \bar{D}_{L_\tau} $ & ($ \omega^3$, 1) & $ \tau_{R} $ & ($ \omega^2$, 1) & $ \nu_{\tau R} $ & (1, -1) & &\\
\hline
\end{tabular}
\begin{center}
\caption{The fields descriptions are same as given in Table(\ref{field_trans}).}
\label{sym_tautau}
\end{center}
\end{small}
\end{table}

Similarly, one can assign the various fields transformation under the action of
($ Z_{8} \times Z_{2} $) to obtain the texture with $ m_{\tau \tau} =0$. 
The form of $M^{(18)}_D$, $M_R$ and $M_S$ used to get $ m_{\tau \tau}  = 0$ are given
in Eq.(\ref{eq:md18_mtautau}). We summarize the fields transformations in the Table \ref{sym_tautau}.
 Here, no extra scalar  singlet is needed to obtain the mass structure of $ M_R $ which has  $ L_e - L_{\mu} $ symmetry 
and for $ M_S $ we need two scalar singlets ($ \lambda_1 $, $ \lambda_2 $) which transform under $Z_8 \times Z_2 $ as 
($ \omega^{2} $, 1) and ($ \omega^{7} $, -1) respectively.
 We also give transformation to singlet field S as ($ \omega $, 1)  under ($ Z_8 \times Z_2 $) which will prevent the term $ \overline{S^{c}} S $. Note that symmetry realization of this texture is more economical than the  $m_{e\tau} = 0$ texture.

\section{Conclusions}
\label{sec7}
In this paper we have studied the low energy phenomenology of the  
minimal extended type I
seesaw model which can 
accommodate an  eV scale light sterile neutrino \cite{Barry:2011wb, Zhang:2011vh}. 
This  model is motivated by the recent experimental
evidences which support the existence of light sterile neutrinos 
in addition to three active neutrinos.
In this model, apart from three right handed neutrinos, an
extra gauge singlet $S$ is added to the SM. Under the minimal extended
seesaw mechanism, this model give rise to three active neutrinos in the 
sub-eV scale with
one of the active neutrinos having vanishing mass and 
one sterile neutrino in the eV scale. 
In this model the Dirac mass matrix, $M_D$, is an arbitrary
$3 \times 3$ complex matrix, the Majorana mass matrix $M_R$
is a $3 \times 3$ complex symmetric matrix and $M_S$ which couples
the right handed neutrinos
and the singlet $S$ is a $1 \times 3$ matrix. 

We obtain different  textures of $M_D$,
$M_R$ and $M_S$ that give rise to phenomenologically
allowed zero textures in the low energy neutrino mass matrix, $m_\nu$. 
The maximum number of zeros in $M_D$
that results in viable $m_{\nu}$ are found to be five. 
Thus, there are 126 different possible
structures of $M_D$ to be probed. 
We consider four possible structures of $M_R$ with  one
diagonal and three non diagonal forms. 
The maximum number of zeros in $M_S$ 
is one as two zeros do not result
in phenomenologically viable textures of $m_{\nu}$. 
This leads to three possible structures of $M_S$. 
After analyzing all the different combinations
we obtain only two viable one
zero textures of $m_\nu$ ($m_{e \tau} = 0$ and $m_{\tau \tau}=0$)
with different possible structures of $M_D$, $M_R$ and $M_S$.
We study these textures of $m_\nu$ in the light of the current
oscillation data. 
Both these textures have inverted hierarchical mass spectrum
and we get constraints on observables like effective 
Majorana neutrino mass $m_{ee}$ and Dirac
CP phase $\delta_{13}$. 
For the texture $m_{e\tau}=0$, we obtain the allowed values of Dirac CP phase
$ \delta_{13} $ is around $ \pm 90^\circ $. 
Note that $\delta_{13} \sim -90^\circ$
is favored by current neutrino oscillation experiments. For $m_{\tau \tau}=0$, 
$\delta_{13}$ lies between (150$^\circ$--240$^\circ$). 
The allowed range for the  effective Majorana mass is
different for both these textures. 
It can thus be used to distinguish between the two textures.
Also, in our study we observed that due to improved constraints on the  
mass squared differences and $\theta_{13}$ the texture $m_{\tau \tau}=0$ 
disfavours higher octant of $\theta_{23}$.

Next we studied the predictions of the MES model for the Yukawa matrices that 
gave viable forms of $m_\nu$  and  check whether  
any extra correlations  can come from the model. 
This is expected since in the framework of this model both the 
active and sterile neutrino masses as well as the active 
sterile mixing depend on the  parameters of the Yukawa matrices 
$M_D$, $M_R$ and $M_S$.  Thus, there may be additional relations
between different observables,  
which are the  predictions of the model. 
We find that some of the 
Yukawa matrices which can generate allowed one zero textures 
$m_{e \tau} = 0$ and $m_{\tau \tau}=0$ in the active neutrino 
mass matrix, $m_\nu$, cannot satisfy the extra correlations 
coming from the predictions of the MES model.  
Our analysis reveals that due to these additional correlations 
among the $126 \times 4 \times 3=1512$ possible combinations of
$M_D$, $M_R$ and $M_S$, only 6 combinations giving
$m_{e\tau}=0$ and other 6 combinations giving 
$m_{\tau \tau}=0$ are allowed from the current oscillation data.
The 6 allowed combinations which give $m_{e\tau}=0$, 
reveal  severe restrictions on the values of $\theta_{23}$ 
and $\theta_{13}$ due to the extra correlations in the MES model 
and only the lower octant of 
$\theta_{23}$ and  relatively higher values of
$\theta_{13}$ remains allowed.
In addition an interesting correlation is obtained connecting the 
mass of the sterile neutrino to the active sterile mixing parameters 
which also involves the light neutrino masses and mixing. 
Thus this correlation connects the active and the sterile sector. 
For $m_{e \tau} =0$ the prediction for the sterile neutrino mass obtained 
from the MES model is 
in complete agreement with what is obtained from global analysis. 
The texture, $m_{\tau \tau}=0$ also
predicts a correlation for sterile neutrino mass. 
This however is in marginal agreement with the  global analysis.
We also explored the consequences of NLO correction terms in our analysis and depicted the parameter space in $ M_D $, $ M_R $ and $ M_S $ for which the NLO corrections can be neglected as compared to the leading order term.
Finally, working within the framework of MES mechanism,
we present simple discrete Abelian symmetry models $Z_8 \times Z_2$ leading to
the two
phenomenologically allowed zero textures of $m_{\nu}$. 



In conclusion, we analyzed the low energy prediction of the minimal extended 
seesaw model that can give an eV scale sterile neutrino. 
We emphasize that this task is performed for the first time in this paper. 
The results described
in our analysis shows the compatibility of this model to the 
neutrino oscillation data.
We also find correlations that can be tested in future experiments.  
This kind of study is indispensable to test the viability of a given model 
in the context of 
present and forthcoming neutrino oscillation experiments.


\acknowledgments
Authors are grateful to Anjan Joshipura for discussions and useful 
comments in the initial stages of the work.
The work of SG is supported by the Australian Research 
Council through the ARC Center of Excellence for Particle Physics (CoEPP Adelaide) at the Terascale (CE110001004).
The work of MG is supported by the ``Grant-in-Aid for Scientific Research of the Ministry of Education,
 Science and Culture, Japan", under Grant No. 25105009.

\bibliography{one_zero.bib}{}

\end{document}